\documentclass[fleqn,usenatbib]{mnras}

\usepackage{newtxtext,newtxmath}
\usepackage[T1]{fontenc}
\usepackage{ae,aecompl}
\usepackage{graphicx}   % Including figure files
\usepackage{amsmath}    % Advanced maths commands
\usepackage{amssymb}    % Extra maths symbols

\usepackage{epsfig}
\usepackage{float}
\usepackage[utf8]{inputenc}
\usepackage{epstopdf}
\usepackage{natbib}
\usepackage{color}
\usepackage{aas_macros}
\usepackage{epstopdf}
\usepackage{amsfonts}
\usepackage{threeparttable}
\usepackage{gensymb}
\usepackage{kantlipsum}
\usepackage{etoolbox}
\usepackage{color}
\usepackage{tikz}
\usepackage{tabularx}
\usepackage{booktabs}
\usepackage[normalem]{ulem}

\newcolumntype{L}[1]{>{\raggedright\let\newline\\
\arraybackslash\hspace{0pt}}m{#1}}
\newcolumntype{C}[1]{>{\centering\let\newline\\
\arraybackslash\hspace{0pt}}m{#1}}
\newcolumntype{R}[1]{>{\raggedleft\let\newline\\
\arraybackslash\hspace{0pt}}m{#1}}

\def\sige{\mbox{$\sigma_{\rm e}$}}

\def\Msun{\mbox{$M_\odot$}}
\def\Mtot{\mbox{$M_{\rm tot}$}}
\def\Lsun{\mbox{$L_\odot$}}
\def\Ysun{\mbox{$\Upsilon_\odot$}}
\def\ML{\mbox{$M/L$}}

\def\Yst{\mbox{$\Upsilon_*$}}
\def\Ystv{\mbox{$\Upsilon_{*}^{\rm var}$}}

\def\Mdyn{\mbox{$M_{\rm dyn}$}}
\def\Ytot{\mbox{$\Upsilon_{\rm tot}$}}

\def\mst{\mbox{$M_{*}$}}

\def\Mvir{\mbox{$M_{\rm vir}$}}
\def\cvir{\mbox{$c_{\rm vir}$}}

\def\fdm{\mbox{$f_{\rm DM}$}}

\def\lsim{\mathrel{\rlap{\lower3.5pt\hbox{\hskip0.5pt$\sim$}}
    \raise0.5pt\hbox{$<$}}}                % less than or approx. symbol
\def\gsim{~\rlap{$>$}{\lower 1.0ex\hbox{$\sim$}}}

\def\sigAp{\mbox{$\sigma_{\rm Ap}$}}

\def\Yssp{\mbox{$\Upsilon_{SSP}$}}

\def\Fig{\mbox{Figure~}}

\def\Sec{\mbox{Section~}}

\def\Re{\mbox{$R_{\rm e}$}}

\def\Remaj{\mbox{$R_{\rm e, maj}$}}

\newcommand{\HI} {{\rm H}\,{\footnotesize\rm I}}

\def\amw{\mbox{$\alpha_{\rm mw}$}}

\title[DM fraction and mass density slope in galaxies]{The dichotomy of dark matter fraction and total mass density slope of galaxies over five dex in mass}

\author[Tortora C. et al.]{\noindent
C.~Tortora$^{1,2}$\thanks{E-mail: ctortora@arcetri.astro.it},
L.~Posti$^{2,3}$, L.V.E.~Koopmans$^{2}$, N.~R.~Napolitano$^{4,5}$
\\~\\
$^1$ INAF -- Osservatorio Astrofisico di Arcetri, Largo Enrico
Fermi 5, 50125, Firenze, Italy\\
$^2$ Kapteyn Astronomical Institute, University of Groningen, P.O.
Box 800, 9700 AV Groningen, the Netherlands \\
$^3$ Observatoire astronomique de Strasbourg, Universit\'{e} de Strasbourg,
CNRS UMR 7550, 11 rue de l'Universit\'{e}, 67000 Strasbourg, France \\
$^{4}$ School of Physics and Astronomy,  Sun Yat-sen University Zhuhai Campus, 2 Daxue Road,  Tangjia,  Zhuhai,  Guangdong 519082,  P.R. China \\
$^{5}$ INAF -- Osservatorio Astronomico di
Capodimonte, Salita Moiariello, 16, 80131 - Napoli, Italy
}

\date{Accepted XXX. Received YYY; in original form ZZZ}

\pubyear{2019}

\hypersetup{draft}

\begin{document}
\label{firstpage}
\pagerange{\pageref{firstpage}--\pageref{lastpage}}
\maketitle

\begin{abstract}
We analyse the mass density distribution in the centres of
galaxies across five orders of magnitude in mass range. Using
high-quality spiral galaxy rotation curves and infrared photometry
from SPARC, we conduct a systematic study of their central dark
matter fraction (\fdm) and their mass density slope ($\alpha$),
within their effective radius. We show that lower-mass spiral
galaxies are more dark matter dominated and have more shallow mass
density slopes when compared with more massive galaxies, which
have density profiles closer to isothermal. Low-mass ($\mst \lsim
10^{10}\, \rm \Msun$) gas-rich spirals span a wide range of \fdm\
values, but systematically lower than in gas-poor systems of
similar mass. With increasing galaxy mass, the values of \fdm\
decrease and the density profiles steepen. In the most massive
late-type gas-poor galaxies, a possible flattening of these trends
is observed. When comparing these results to massive ($\mst\gsim
10^{10}\Msun$) elliptical galaxies from SPIDER and to dwarf
ellipticals from SMACKED, these trends result to be inverted.
Hence, the values of both \fdm\ and $\alpha$, as a function of
\mst, exhibit a U-shape trend. At a fixed stellar mass, the mass
density profiles in dwarf ellipticals are steeper than in spirals.
These trends can be understood by stellar feedback from a more
prolonged star formation period in spirals, causing a
transformation of the initial steep density cusp to a more shallow
profile via differential feedback efficiency by supernovae, and by
galaxy mergers or AGN feedback in higher-mass galaxies.
\end{abstract}


%%%%%%%%%%%%%%%%%%%%%%%%%%%%%%%%%%%%%%%%%%%%%%%%%%

%%%%%%%%%%%%%%%%% BODY OF PAPER %%%%%%%%%%%%%%%%%%

\section{Introduction}\label{sec:intro}

Dark matter (DM) dominates the mass density of galaxies and
clusters of galaxies. Its budget amounts to $\sim 85$ per cent of
the total mass density of the universe (e.g.,
\citealt{SDSS_DR1,SDSS_DR6,SDSS_DR7_Abazajian}) and its imprint is
found on cosmological scales over the entire history of the
Universe (e.g., \citealt{Komatsu+11_WMAP7}). Within the standard
cosmological framework, i.e. the $\Lambda$CDM model, numerical
simulations of (DM only) structure formation have explained the
formation of virialised DM haloes from tiny initial density
perturbations, constraining the shapes and the properties of DM
haloes. The spherically averaged density profile, $\rho_{\rm DM}
(r)$, of DM haloes, is found to be nearly independent of halo mass
and universal, and is well described by a double power-law profile
with $\rho_{\rm DM} (r) \propto r^{-3}$ in the outer regions and
$\rho_{\rm DM} (r) \propto r^{\alpha}$, with $\alpha < 0$, in the
centre (\citealt{NFW96}, hereafter NFW; \citealt{Bullock+01};
\citealt{Maccio+08}). However, measurements of the rotation
velocities of gas in DM-dominated low-mass spiral galaxies have
cast some reservations on such a universality, since the circular
velocity in these systems is observed to rise linearly with
radius, suggesting density cores rather than cusps ($\alpha \sim
0$, e.g. \citealt{deBlok10}). The \cite{Burkert95} profile is the
prototype of cored models and has been shown to reproduce the DM
profile of late-type galaxies (LTGs; sometimes also referred to as
spiral galaxies) quite well (e.g. \citealt{Salucci_Burkert00}).
Instead, in early-type galaxies (ETGs; i.e., ellipticals and
lenticulars), gravitational lensing and central stellar dynamics
suggest that a cuspy profile is typically preferred
(\citealt{NRT10}; \citealt{Napolitano+11_PNS};
\citealt{Cappellari+13_ATLAS3D_XX}; \citealt{Tortora+10lensing,
TRN13_SPIDER_IMF, Tortora+14_DMslope};
\citealt{Mukherjee19_SEAGLEII}). Whether these differences are due
to some physical process that is not entirely represented in
numerical simulations, or due to a failure of the CDM paradigm, is
still actively debated.

One way to address this problem, and constrain galaxy-formation
models, is to study scaling relations among their DM halo
parameters and stellar quantities. There is increasing evidence
that a critical stellar mass scale around  $\sim 3 \times
10^{10}\, \rm \Msun$ ($\sim 10^{12}\, \rm \Msun$ in virial mass)
exists, corresponding to transitions or even breaks in the trends
of different scaling relations.

If this is indeed constitutes a fundamental mass scale in galaxy structure, then it is quite plausible that also physical processes responsible for galaxy evolution change when crossing this mass scale. Such a characteristic mass is observed in the trends with galaxy mass of the total \ML\ and star formation efficiency (when considering all galaxies, e.g. \citealt{Benson+00}, \citealt{MH02}, \citealt{vdB+07}; \citealt{CW09}; \citealt{Moster+10}; though it appears different when considering galaxies of different types, e.g. \citealt{Dutton+10}; \citealt{More+11}; \citealt{Wojtak_Mamon13}; \citealt{Posti+19}), the half-light dynamical \ML\ (\citealt{Wolf+10}; \citealt{Toloba+11_I}), the central DM fraction (\citealt{Cappellari+13_ATLAS3D_XX}; \citealt{TLBN16_IMF_dwarfs};
\citealt{Lovell+18_Illustris}), the $\mu_e-\Re$
(\citealt{Capaccioli+92a}; \citealt{TullyVerheijen97};  \citealt{Kormendy+09}) and the size-mass (\citealt{Shen+03};  \citealt{HB09_curv}) relations, the trends in optical colour, metallicity and stellar \ML\ gradients (\citealt{Spolaor+10}; \citealt{Kuntschner+10}; \citealt{Tortora+10CG, Tortora+11MtoLgrad}), as well as the gradients in the dynamical \ML\ profiles through several \Re\ (\citealt{Napolitano+05}).

In this paper, we uniformly analyse the stellar and dark matter
distribution in galaxies of different types, providing some of the
most comprehensive constraints on galaxy formation models over
five orders of magnitude in stellar mass. Studies of the DM
fraction and total mass density slope in the central regions of
galaxies have particularly focused In the last years on ETGs, due
to the wealth of dynamical and gravitational-lensing data (e.g.,
\citealt{Cappellari+06}; \citealt{Bolton+06_SLACSI};
\citealt{Bolton+08_SLACSV};
\citealt{Tortora+09,SPIDER-VI,Tortora+14_DMevol,
Tortora+14_DMslope, Tortora+18_KiDS_DMevol};
\citealt{Auger+10_SLACSX}; \citealt{ThomasJ+11};
\citealt{Oguri+14}; \citealt{Dutton_Treu14}). Using similar
observables, we strive at expanding this analysis to a broader
range of galaxy types, investigating the mass density profile in
the central regions of late-type galaxies. In particular, we
concentrate on their central DM fraction and the total mass
density slope, both derived within the effective radius, \Re. We
apply a uniform analysis method, which we have developed in the
past for ETGs and dwarf ellipticals (\citealt{Tortora+09,
SPIDER-VI, Tortora+14_DMevol, Tortora+14_DMslope,
TLBN16_IMF_dwarfs, Tortora+18_KiDS_DMevol}).

The central DM content in massive ETGs is very well studied (e.g.,
\citealt{Gerhard+01}; \citealt{Cappellari+06};
\citealt{ThomasJ+07,ThomasJ+11}; \citealt{Tortora+09}). In
particular,  \cite{TLBN16_IMF_dwarfs} have proposed that the DM
fraction with galaxy mass exhibits a U-shape trend, with large DM
fractions in both the most, and least, massive galaxies (see also
\citealt{Lovell+18_Illustris}). Furthermore, gravitational lensing
and central stellar dynamics suggest that the stellar and DM
profiles conspire to yield a total mass density profile which is
nearly isothermal in massive ETGs (e.g., \citealt{Kochanek91};
\citealt{Bolton+06_SLACSI}; \citealt{Koopmans+06_SLACSIII,
Koopmans+09}; \citealt{Gavazzi+07_SLACSIV};
\citealt{Bolton+08_SLACSV};
\citealt{Auger+09_SLACSIX,Auger+10_SLACSX}; \citealt{Chae+14};
\citealt{Oguri+14}), i.e. having a total mass density profile
following $\rho(r) \propto r^{\alpha}$ with $\alpha \sim -2$ and a
scatter of $\sim 10\%$. However, lower masses ETGs seem to show a
non-universal total mass density slope as it generally steepens at
lower masses (e.g., \citealt{Dutton_Treu14};
\citealt{Tortora+14_DMslope}). In contrast, while the amount of
dark-to-luminous matter density in the centres of spirals has been
extensively studied (e.g. \citealt{Swaters+14};
\citealt{ErrozFerrer+16}; \citealt{Lelli+16_dens}), it is not
straightforward to compare these studies with those of ETGs at
face value.

To homogeneously compare the dark matter fractions of LTG sand ETGs, we have therefore performed an analysis of LTGs HI rotation curve data which is similar to what is usually done for central velocity dispersions of ETGs. While stellar kinematics in early-type galaxies cannot break the stellar-dark-matter degeneracy, except for isolated cases which rely on excellent and spatially extended dynamical data (see e.g. \citealt{Napolitano+14}), for local late-type galaxies with measured extended rotation curves, the mass density profile can be directly inferred, with limited modelling assumptions and degeneracies. The SPARC sample (\citealt{Lelli+16_SPARC}) represents the ideal dataset to perform such a study,
because it combines \HI\ kinematics (which traces the circular velocity) with 3.6 $\mu$m photometry (tracing the old stellar mass distribution). We compare the results in this paper with theoretical expectations and independent observational results for ETGs and dwarf ellipticals, providing an homogeneous and self-consistent picture of galaxy evolution across a wide range of masses and galaxy types.

The paper is organised as follows. In \Sec\ref{sec:data}, we present the galaxy datasets that we use, the DM fraction and the mass density slope derivation. The DM fraction and the mass density slope in terms of stellar mass are presented and discussed in \Sec\ref{sec:density_slope}. In \Sec\ref{sec:phys_interpretation}, we provide a physical interpretation of the results, while our conclusions are given in \Sec\ref{sec:conclusions}. Decimal logarithms are used in the paper. If not stated otherwise, we adopt a cosmological model with $(\Omega_{m},\Omega_{\Lambda},h)=(0.3,0.7,0.75)$, where $h = H_{0}/100 \, \textrm{km} \, \textrm{s}^{-1} \, \textrm{Mpc}^{-1}$ (\citealt{Komatsu+11_WMAP7}).

\section{Data samples and analysis}\label{sec:data}

In this section, we describe the data samples and the analysis adopted to derive the stellar and total mass density profiles. In \Sec\ref{subsec: SPARCS}, we start with local spiral galaxies from the SPARC sample, which span a stellar mass range from $\sim 10^{7}$ to $\sim 10^{11}$ \Msun. In \Sec\ref{subsec:SPIDER}, we introduce ETGs from the SPIDER sample, while their lower-mass counterparts from the SMACKED sample, i.e. dwarf ellipticals (dE), are presented in \Sec\ref{subsec:SMACKED}.

\subsection{Late-type Galaxies from the SPARC sample}\label{subsec: SPARCS}

We start from the sample of 175 galaxies from the SPARC database (\citealt{Lelli+16_SPARC} for more details) with extended \HI\ rotation curves and Spitzer [3.6] photometry. Although SPARC is neither a statistically complete nor a volume-limited sample, it is representative of disk galaxies in the nearby Universe. SPARC spans a wide range in morphologies (S0 to Im/BCD), stellar masses ($\sim 10^{7}$ to $\sim 10^{11}$ \Msun), effective radii ($\sim$0.3 to $\sim$15 kpc), rotation velocities ($\sim$20 to
$\sim$300 km~s$^{-1}$), and gas content ($0.01 \lsim
M_{\HI}/L_{[3.6]}/(\Msun/\Lsun)\lsim 10$). Throughout, we define the effective radius as the radius encompassing half of the total [3.6] luminosity. \cite{Lelli+16_SPARC} performed a simple photometric bulge plus disk decomposition on the sample. They find that 32 out of the original 175 galaxies have a non-negligible bulge component and concentrate at very high luminosities and low gas-mass fractions. The total luminosity is converted to a total stellar mass assuming a [3.6] stellar mass-to-light ratio, \Yst, of $0.6 \Ysun$ \footnote{Following \cite{Lelli+16_Tully-Fisher},
we assume that \Yst\ is almost constant in the [3.6] band.
Although a consensus on the overall normalisation has not
been reached, \cite{Lelli+16_Tully-Fisher} find that a value
$\gsim \, 0.5$ minimises the scatter in the Tully-Fisher relation,
consistently with what is expected in a $\Lambda$CDM cosmology. Moreover, for the disc and bulge components, stellar population synthesis models suggest the following values: $\Upsilon_{\rm bulge} = 0.7$ and $\Upsilon_{\rm disc} = 0.5$ (e.g. \citealt{Schombert_McGaugh14}). Using these results, we assume a nominal value of $\Yst = 0.6 \, \Ysun$ for the stellar \ML. We notice that the specific value of \Yst\ does not affect the calculation of total mass density slopes, but it does impact stellar mass and DM fraction calculations. We will discuss the effect of this assumption on our conclusion later in the paper and demonstrate that it will be almost negligible.}. Distances to these galaxies are
measured in various ways. The best distance measurements, however, are determined from the tip of the red giant branch, the Cepheids Magnitude-Period Relation, the Supernovae light curves, and using the distance of the cluster for galaxies in the Ursa Major Cluster. The typical distance errors are $\sim 5$ to $10$ per cent, but can reach uncertainties up top $30$ per cent for distances derived from the Hubble-Flow. These latter assume $H_{0}=73 \, \rm km s^{-1} Mpc^{-1}$ and are corrected for Virgo-centric infall.

\cite{Lelli+16_SPARC} have derived rotation curves from literature data, mainly based on \HI\ data. However, for some galaxies, hybrid rotation curves, which combine \HI\ and $H_{\alpha}$ measurements, are used. In what follows, we will consider only stars and neutral hydrogen in the baryonic mass budget, neglecting molecular gas, which should be dynamically unimportant in most circumstances
(e.g. \citealt{Saintonge+11_COLDGASS_I}).

Out of the 175 galaxies in the SPARC sample, we consider only those with inclinations larger than $30^{\circ}$, because the rotation velocities for nearly face-on systems are highly uncertain. This selection does not introduce any bias in the sample selection since galaxies are randomly oriented on the sky. We also cut those systems for which \Re\ is not covered by the rotation curve out of our final sample, in order to avoid extrapolations of the inferred rotation curve and mass distribution. We are then left with 152 out of 175 galaxies.

The deprojected mass profile $M(r)$ is determined by assuming $M(r) = V^{2} r/G$, where $V$ is the intrinsic azimuthal velocity that is obtained after deprojecting the measured velocity on the sky. The possibility to use this approximation, despite galaxies are not fully spherical, is based on the following arguments:
\begin{itemize}
\item[(i)] The above formula holds for an exponential disk (within 15 per cent, see \citealt{Binney_Tremaine08}, S2.6, Fig.~2.17), for a flattened spheroidal distribution (within 10 per cent, see \citealt{Binney_Tremaine08}, S2.5, Fig.~2.13),  and is rigorously valid for the Mestel disk model.
\item[(ii)] Since observationally we have no indications on the geometry of the DM haloes of LTGs, computing the total matter distribution with the above formula is a reasonable assumption.
\item[(iii)] Considering that the SPARC galaxies have been selected to have regular kinematics and minimal levels of non-circular motion, this mass inference should hold to good accuracy. We neglect the velocity dispersion of \HI, which has a typical value of $\sim 8 \, \rm km/s$ and yields a correction of $\sim 10$ per cent to the velocity in most of the cases.
\end{itemize}

The total mass profile is determined by linear interpolation of the data points\footnote{We have checked that using different interpolating functions (e.g. polynomial functions of a different degree) negligibly affects our results.}. A more complex analysis is necessary for the mass density slope, due to the discrete measurement
of the rotation curves. To avoid artefacts, we interpolate the
rotation curves with polynomials. We carry out a weighted fit with a 4th order polynomial of the ten data-points closest to \Re\ in the observed rotation curves. We have visually inspected both the rotation curves and mass profiles, to assess the quality of the fit. All rotation curves appear well fitted. We have also verified that changing the number of points that are fitted or the order of the polynomial
does not qualitatively affect our conclusions.

To determine the errors on \mst, we use the formula in
\cite{Lelli+16_Tully-Fisher}, propagating the errors on the  distances, luminosities and stellar \ML\ values. For the effective radii, we adopt an average error of $0.2$ dex\footnote{Unfortunately, we do not have accurate estimates for the errors on the effective radius, except for the contribution from the typical errors on distances ($\sim 5-10$ per cent). However, we assume a conservative value of 0.2 dex.
The exact value of this error component will not affect our conclusions.}. Finally, to calculate the errors on the DM fraction and mass density slope, we create a set of 1,000 Monte Carlo realisations of the velocity profile $V(r)$, assuming Gaussian errors $\delta V$ (which mainly account for differences in the approaching and receding side of the \HI\ rotation curve). We calculate the dark matter fraction and mass density slope for each realisation. The errors
are subsequently defined  as the standard deviation of the resulting distributions. We find an error of 20 and 16 per cent, respectively, on the dark matter fraction and mass density slopes.

\subsection{ETGs and dEs}\label{subsec:ETGs_dEs}

To complement the LTG analysis, here we introduce
two samples of early-type systems: massive ETGs and dEs.

\subsubsection{ETGs: SPIDER sample}\label{subsec:SPIDER}

For massive ETGs, we use the local ($0.05<z<0.095$) sample of
$\sim 4300$ giant ETGs drawn from the complete SPIDER survey (see
\citealt{SPIDER-I} and \citealt{SPIDER-VI} for further details
about the sample selection). The SPIDER dataset includes stellar
masses derived from fitting stellar population synthesis (SPS)
models to their optical and near-infrared photometry
(\citealt{SPIDER-V}) using a \cite{Chabrier01} Initial Mass
Function (IMF). It also includes galaxy structural parameters
(effective radius \Re\ and S\'ersic index $n$; using 2DPHOT,
\citealt{LaBarbera_08_2DPHOT}), homogeneously derived from $g$
through $K$ wavebands, and the SDSS central-aperture velocity
dispersions, $\sigAp$, within a circular fibre aperture of $1.5''$
radius. SPIDER ETGs are defined as luminous bulge-dominated
systems, featuring passive spectra in the central SDSS fibre
aperture (\citealt{SPIDER-I}).

\subsubsection{dEs: SMACKED sample}\label{subsec:SMACKED}

At masses lower than $10^{10} \, \Msun$, we use the
dwarf ellipticals (dEs) from \cite{TLBN16_IMF_dwarfs}. We analyze
the sample of 39 dEs in the magnitude range $-19 < \rm M_{\rm r} <
-16$, selected from the Virgo Cluster Catalog (VCC,
\citealt{Binggeli+85}). Albeit incomplete in luminosity, this
sample is representative of the early-type population in this
magnitude range (\citealt{Toloba+14_II}). The H-band structural
parameters (the major-axis effective radius, \Remaj, S\'ersic
index, n, and axis ratio, q) are taken from \cite{Toloba+14_II}
and \cite{Janz+14}. For 9 systems without a measured value of $n$
(as they had no fit with a single S\'ersic component or are not
present in \citealt{Janz+14}), we adopted $n=1$. The effective
velocity dispersions, \sige, computed within an ellipse of
semi-major axis length \Remaj\ are used (\citealt{Toloba+14_II}).
We obtain the stellar H-band mass-to-light (\ML)
ratio, \Yssp, for each galaxy, using the best-fit age and metallicity from
\cite{Toloba+14_II}, and the simple stellar population (SSP)
models of~\cite{Vazdekis+12}, for a Kroupa IMF. These \Yssp\ values are
converted to those for a Chabrier IMF by subtracting $0.05$ dex
(i.e. the difference in normalisation between the Kroupa and
Chabrier IMFs; \citealt{Tortora+09}).

\subsubsection{Model assumptions and mass modelling}

According to \cite{ML05a,Tortora+09,SPIDER-VI,TLBN16_IMF_dwarfs}, we model
the aperture velocity dispersion of individual galaxies using the
spherical isotropic Jeans equations to estimate the (total)
dynamical mass \Mdyn\ (which, we will also refer to as total
mass $\Mtot$, since it includes all the mass from all the
components: stars, gas and DM). In the Jeans equations, the
stellar mass density and the total mass distribution need to be
specified. The stellar mass density is provided by the
deprojection of the S\'ersic fit of the $K$-band and $H$-band
galaxy images, for SPIDER and SMACKED samples, respectively. In the following we will present the mass models adopted for the DM or total mass distribution.

\begin{itemize}
\item {\it Reference NFW model with a non-universal IMF.} As a reference model, we assume a two-component model, composed of an
NFW profile for the DM (motivated by N-body simulations) and
a deprojected S\'ersic profile for the stellar mass with a constant
stellar \ML. This model is parameterised by the virial
concentration index \cvir\ and the (total) virial mass \Mvir\
(\citealt{NFW96, NFW97}). We fix the DM halo parameters using the
correlation between \Mvir\ and \cvir, from N-body simulations
based on WMAP5 cosmology \citep{Maccio+08}, as well as the
\Mvir--$\mst$ correlation from abundance matching results in
\citet{Moster+10}, which assumes a Chabrier IMF for \mst. For each
galaxy with a Chabrier stellar mass \mst, the values of \Mvir\ and
\cvir\ and the DM profile are fully determined. The stellar mass
derived from SPS is only used to link each galaxy to the correct
halo, using the correlations mentioned above. In the stellar
profile the stellar \ML, \Ystv, is free to vary. These results are
taken from \cite{TRN13_SPIDER_IMF} and \cite{Tortora+14_DMslope}
for the SPIDER sample and from \cite{TLBN16_IMF_dwarfs} for the
SMACKED sample.

Alternatives to the standard NFW profile could also be
considered. In particular, the NFW profile could be steeper (due
to contraction by the baryonic component; \citealt{Gnedin+04}).
For massive ETGs, \cite{Tortora+14_DMslope} have shown that this
introduces a small effect in the total mass density slopes,
producing a slightly shallower trend with \mst\ (of $\sim 5\%$ in the less massive ETGs), but increases the
DM fractions (\citealt{TRN13_SPIDER_IMF}). We have also analysed the impact of fixing the virial mass to, e.g. a unrealistic constant value of $10^{13}\, \rm \Msun$, finding a slightly shallower trend with \mst. Larger changes are
induced if a Burkert profile or a high-concentration NFW model are adopted. In the latter case, very shallow average mass density slopes are found ($\amw \sim -1.6$ at $\mst \sim 3 \times 10^{11}\, \rm \Msun$), which do not match the results from strong lensing analysis (e.g, \citealt{Koopmans+09}). Instead, fixing the IMF
to the standard Chabrier one in massive and high-velocity dispersion galaxies is in contrast with different results pointing to a bottom-heavy IMF (e.g.; \citealt{Cappellari+12}; \citealt{Spiniello+12}; \citealt{TRN13_SPIDER_IMF}). For dEs, a
systematic analysis of different model assumptions has been made
in \cite{TLBN16_IMF_dwarfs}. Since the NFW model provides a fairly
good and homogeneous approximation of the DM distribution in ETGs
and dEs, we will adopt this model assumption in the rest of the
paper, except if otherwise stated.

\item {\it Alternative models with a universal IMF.} In \Sec\ref{subsec:comparison_simulations}, we will compare our
results with simulations, which assume a universal IMF.
Indeed, we also use alternative mass profiles for both the
samples, which model the total mass distribution. We assume that
the mass follows the light, as $M_{\rm const-M/L}(r)= \Ytot L(r)$,
where $L(r)$ is the deprojected luminosity of the S\'ersic profile, \Ytot\ the only free parameter of the model and we set the IMF to the Chabrier one. For the sample of
massive ETGs, we also explore the case of an isothermal mass
profile, which is suggested by strong lensing analyses (e.g.
\citealt{Koopmans+09}). We assume a Singular Isothermal Sphere
(SIS) with $M_{\rm SIS}(r) \propto \sigma_{\rm SIS}^{2} r$ and
$\sigma_{\rm SIS}$ being the free parameter. More information can
be found in \cite{SPIDER-VI} and \cite{TLBN16_IMF_dwarfs}, for
ETGs and dEs, respectively.
\end{itemize}

After the mass model is chosen and the predicted velocity
dispersion, $\sigAp^{J}(p)$ is derived from the Jeans equation,
the equation $\sigAp^{J}(p) = \sigAp$ is solved with respect to
the free parameter $p$. The parameter $p$ is equal to \Ystv, \Ytot\
and $\sigma_{\rm SIS}$ for the three models discussed above. For
the SPIDER galaxies, the velocity dispersions are defined within
the circular aperture of the SDSS fibre. Instead, for dEs we
calculate the 3D velocity dispersion from the radial Jeans
equation at the circularised (geometric) effective radius, to
account for the fact that \sige\ is averaged within an elliptic
aperture, while we rely on spherical models.

\begin{figure*}
\psfig{file=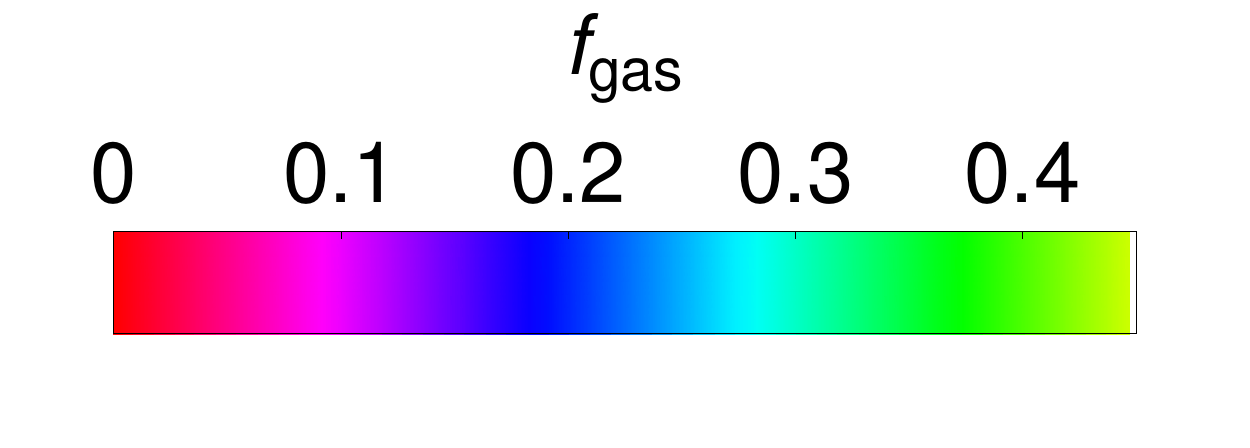, width=0.25\textwidth}
\psfig{file=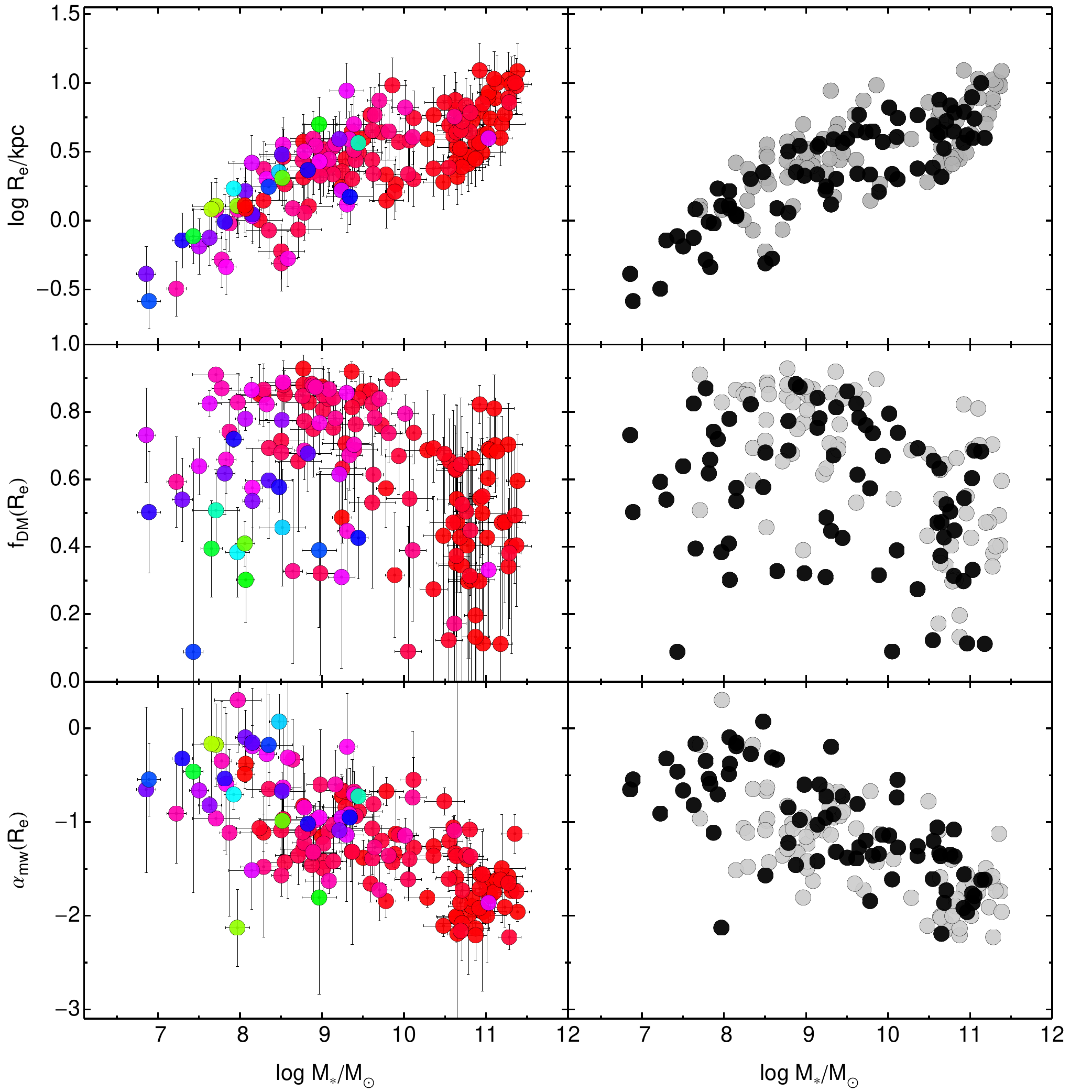,
width=0.8\textwidth} \caption{Effective radius \Re\ (top panels),
DM fraction within 1 \Re\ \fdm\ (middle panels), and mass density
slope \amw\ (bottom panels) are plotted as a function of
stellar mass, \mst, for the SPARC sample. Error
bars for \Re\ are fixed to $0.2$ dex, the errors for the other
quantities are determined as described in the main text. {\it
Left.} The points are colour-coded according to the gas fraction
within \Re, $f_{\rm gas}= M_{\rm gas}(\Re) / M_{\rm tot}(\Re)$
(gas decreases from green, passing through blue, till to the gas
poorest in red), a coloured bar is added on the top of the figure.
We omit the dependence on the galaxy type, since it is providing
similar changes of $f_{\rm gas}$. {\it Right.} With black (grey)
symbols we show the galaxies with more (less) accurate distance
measurements
(\citealt{Lelli+16_SPARC}).}\label{fig:Re_fdm_slope_Mstar_LTGs_points}
\end{figure*}

\section{Dark matter fraction and mass density slope}\label{sec:density_slope}

We define the 3D de-projected DM fraction within a radius $r$, as
$\fdm (r) = 1 - M_{\rm b}(r) / \Mtot(r)$, where $M_{\rm b}(r)$ and
$\Mtot(r)$ are the baryonic (stars and gas) and total mass as a
function of the de-projected radius $r$. The latter includes
baryons and DM (\citealt{Tortora+09}; \citealt{Auger+10_SLACSX}).
For LTGs gas provides a non-negligible contribution
to the mass budget, while it is negligible in ETGs and dEs
(\citealt{Courteau+14_review}; \citealt{Li+17_IMF}). We
also define the {\it mass-weighted logarithmic density slope},
\amw, within a given radius $r$ (\citealt{Koopmans+09};
\citealt{Dutton_Treu14}; \citealt{Tortora+14_DMslope}) as:
\begin{equation}
\amw(r) = - 3 + d\log \Mtot(r)/d\log r.
\end{equation}
The value $\amw = -2$ corresponds to a total mass density
following an isothermal profile. We will calculate both $\fdm(r)$ and
$\amw(r)$ at the 2D projected effective radius, \Re, and in
what follows we refer to them simply as \fdm\ and \amw\, for
the sake of brevity.

\subsection{Dark matter and mass density slope in SPARC LTGs}

\Fig\ref{fig:Re_fdm_slope_Mstar_LTGs_points} shows the effective
radius, \Re, DM fraction within \Re, \fdm, the mass-weighted
slope at \Re, \amw, as a function of total stellar mass, \mst.
Data-points are colour-coded, in the left panels, according to the
gas fraction within \Re. This fraction is defined as the ratio of
gas and total mass within \Re, $f_{\rm gas} = M_{\rm
gas}(\Re)/\Mtot(\Re)$. We have verified that the impact of the gas
on the central regions is negligible at $\mst \gsim 10^{10}\, \rm
\Msun$ and that, averaging across the sample, $f_{\rm
gas}(\Re) \sim 4$ per cent (median). About 83 per cent of the galaxies have
$f_{\rm gas}(\Re) < 10$ per cent. While the gas content is accounted for
in the DM calculation, the mass density slope is calculated from
the total mass profile\footnote{However, we have verified
that the impact of possible systematic uncertainties in the gas
content on the average slopes is negligible.}.

The effective radius in spiral galaxies is positively
correlated with stellar mass (\citealt{Courteau+07};
\citealt{Mosleh+13}; \citealt{Lange+15}; \citealt{Roy+18}),
similarly to ETGs. The lowest-mass spirals, which are also
systematically gas-richer and have later morphological Hubble
types, have $\Re \sim 0.3 \, \rm kpc$. The most massive spirals,
typically classified as S0/Sa have $\Re \sim 10 \, \rm kpc$,
values similar to the sizes of massive ellipticals (see
\Sec\ref{subsec:comparison_ETGs_dEs}). For masses $\gsim 10^{11}\,
\rm \Msun$ the \Re--\mst\ trend is dominated by gas poorer LTGs
and is steeper. This trend resembles the steep \Re--\mst\
correlation found in massive ETGs, with the only caveat that the SPARC sample lacks galaxies at $\mst\sim 10^{10}\Msun$, which is precisely the transition region where the trend appears to steepen. Overall, the \Re--\mst\ correlation is
statistically significant at more than 99 per cent confidence
level. We fit the relation $\Re \propto \mst^{\gamma}$ and find a
slope value of $\gamma = 0.23 \pm 0.02$.

The main results are shown in the middle and bottom panels of
\Fig\ref{fig:Re_fdm_slope_Mstar_LTGs_points}.
We first show the central DM fraction within one effective
radius, \fdm, as a function of stellar mass. Spirals less massive
than $\sim 10^{10}\, \rm \Msun$ are more DM dominated than the
most massive galaxies. We fit the relation $\fdm \propto
\mst^{\gamma}$, finding $\gamma = -0.056 \pm 0.012$, the
correlation is mild but significant at $>99$ per cent. Among the
galaxies with $\mst \lsim 10^{10}\, \rm \Msun$, the gas-poorest
ones (with $f_{\rm gas} < 5$ per cent) have the largest DM fractions,
i.e. on average $0.81_{-0.18}^{+0.07}$,
where median and 16-84th quantiles of the sample distribution
are quoted. Instead, the gas richer systems with $f_{\rm gas}
\geq 5$ per cent have lower \fdm\ values and a wider distribution, with a
median of $0.65_{-0.24}^{+0.21}$. The most massive spirals,
with $\mst \sim 10^{11}\, \rm \Msun$, have DM fractions
distributed in the whole range 0.1-0.8, with a median of
$0.48_{-0.14}^{+0.21}$ if we only consider the galaxies with $\mst > 10^{11}\, \rm \Msun$. Finally, in the bottom panel, we plot the mass
density slope \amw\ and find that it is inversely correlated
with stellar mass, i.e. it is more negative at larger masses.
Similarly to the previous correlations, this is also
significant at more than 99 per cent and the best-fit slope
of a linear relation of the type $\amw = A + \gamma \log \mst$
is found to be $\gamma = -0.35 \pm 0.03$.
This means that the total mass profile of spiral galaxies is
getting steeper and steeper with mass. A similar regularity was
already noticed by \cite{Lelli+13}, who find a tight correlation between the circular-velocity gradient in the innermost regions of galaxies and their central surface brightness.
The lowest mass and gas-richest systems with $\mst \sim 10^{7}\,
\rm \Msun$ have the shallowest central slopes (i.e. $\amw \sim
-0.5$ on average). Instead, the most massive (gas-poor) spirals
have steeper slopes, approaching the isothermal value at the
largest masses\footnote{Note that these massive LTGs are
already DM-dominated at the effective radius, where the rotation
curve is rather flat, which means that the total mass profile is
isothermal.}.

Instead, in the right panels of
\Fig\ref{fig:Re_fdm_slope_Mstar_LTGs_points} we analyse a possible
source of systematics which can come from the sample selection. In
black, we plot the 73 galaxies with the best distance measurements,
and we show in grey the 79 galaxies with the less accurate
Hubble-Flow distances. We notice that the scatter and the average
trends are not considerably affected by larger errors on
distances. For this reason, we proceed with the whole sample of
152 galaxies.

\begin{figure}
\psfig{file=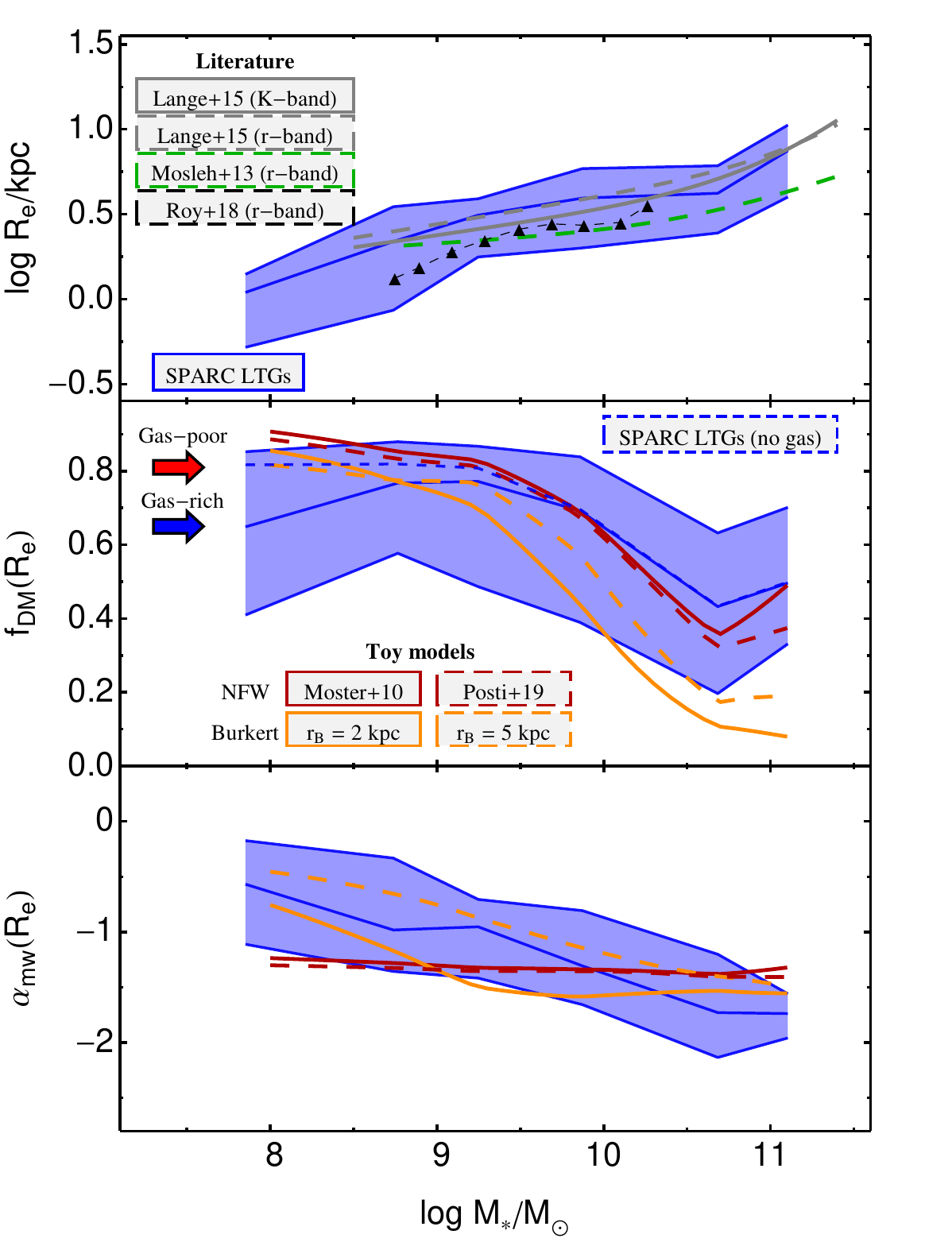, width=0.49\textwidth}
\caption{Effective radius, \Re, DM fraction within 1 \Re,
\fdm\ and mass density slope \amw\ as a function of stellar
mass, \mst, for the SPARC sample. Blue lines and shaded regions
represent the median and 16-84th percentiles in mass bins for
SPARC sample. In the top panel the \Re--\mst\ relation is compared
with some literature (see legend, see \citealt{Roy+18} for a detailed description of the plotted results from the literature). In the middle and bottom
panels, red (orange) lines are the expectations from the NFW
(Burkert) + baryons toy-models, listed in the legend. In the
middle panel, the median value of \fdm\ for gas-rich and gas-poor
low-mass galaxies are also shown. See the text for more
details.}\label{fig:Re_fdm_slope_Mstar_LTGs}
\end{figure}

In \Fig\ref{fig:Re_fdm_slope_Mstar_LTGs} the same results in
\Fig\ref{fig:Re_fdm_slope_Mstar_LTGs_points} are shown as shaded
regions, which represent the median and 16-84th percentiles in
mass bins. In the top panel, the average size-mass relation is plotted, and
compared with some literature data. In particular, we compare with
the best-fit relation in \cite{Mosleh+13} (late-type galaxies in
Table~1), \cite{Lange+15} (morphologically selected late-type
galaxies in Table~2) and \cite{Roy+18} (blue and disk-dominated galaxies), which measured \Re\ in r-band. \cite{Mosleh+13} and \cite{Lange+15} use major axis effective radii, instead \cite{Roy+18} adopt circularised radii. We
also plot the K-band \Re\ from \cite{Lange+15}, which is closer
to our [3.6] effective radius.

\subsection{Comparison to toy models}

In the middle and bottom panels of \Fig\ref{fig:Re_fdm_slope_Mstar_LTGs},
we compare the median \fdm\ and \amw\ (plotted as blue lines and shaded regions) with the expectations
from a set of toy models. For completeness we also show, as dashed blue line,
the \fdm--\mst\ trend when the \HI\ component is neglected. The effect is clearly important only at low masses, where the non-null \HI\ gas mass decreases the DM content. Both \fdm\ and \amw\ are derived directly
from the observed velocities, without any assumption on the
mass model, thus the comparison with specified DM distributions
can be interesting and instructive. The toy models are based on
our reference NFW and Burkert models, by computing the stellar
mass model according to the exponential profile\footnote{
As already discussed before, \cite{Lelli+16_SPARC}
have performed a photometric bulge+disk decomposition of these
galaxies, finding that only 32 out of the original 175 galaxies
have a non-null bulge component (27 out of 152 galaxies discussed
here). These are very few galaxies and assuming also for them that
a single S\'{e}rsic component can approximate their light distribution
negligibly impacts our trends.} (i.e. assuming S\'ersic index $n=1$)
with a Chabrier IMF, and adopting the average size-mass relation of
the SPARC galaxies shown in the top panel of the same figure.
The model predictions do not take into account the gas content, assuming that the small fraction of \HI\ gas is adsorbed in the DM component. This assumption does slightly impacts the observed \fdm\ trend at low masses (dashed vs. solid blue lines in the midlle panel of \Fig\ref{fig:Re_fdm_slope_Mstar_LTGs}). We make very simplistic assumptions, without pretending to
determine the best combination of parameters reproducing both \fdm\
and \amw\ trends. We embed the galaxies in NFW haloes assuming the
\cvir--\Mvir\ and \Mvir--\mst\ correlations used for modelling ETGs
and dEs (\Sec\ref{subsec:ETGs_dEs}). In the Burkert model, the density
and scale parameter ($\rho_{\rm B}$ and $r_{\rm B}$, respectively) are
assumed to follow the relation from \cite{Salucci_Burkert00}.

The expectations for the NFW profile (plotted as a red line)
reproduce quite well the trend of \fdm\ with mass, almost
perfectly overlapping with the observed trend at $\mst \gsim 3
\times 10^{9}\, \rm \Msun$. At lower masses, the toy-model is still in very good agreement with the observed median trend, especially when not considering the gas component (dashed blue line) and for gas-poor systems. The lower \fdm\ of gas-rich
galaxies can be matched using smaller \Mvir\ values than those
predicted by the Moster relation, implying lower star formation
efficiencies. On average, the total mass density slope predicted
using the NFW toy-models is fairly constant with mass across
the whole mass range and not too far from where the observed
\amw\ lie, but it does not reproduce the steepening of the mass
density slope with \mst. Toy-model DM slopes, $\alpha_{\rm DM}$, are on average $\sim -1.2$, consistent with the best-fitting models in \cite{Posti+19}.

The models assuming Burkert profiles (with $r_{\rm B}$ values of 2
or 5 kpc) resemble quite well the observed \fdm--\mst\ trend.
These models are in better agreement with lower-mass spirals, but
tend to have less DM than observed at $\mst \gsim \, 3 \times 10^{9}\, \rm \Msun$.
The average normalization of the mass density slopes and the
observed steepening with mass are, instead, reproduced quite well.
This possibly indicates that spiral galaxies seem to statistically
prefer a Burkert profile for the DM distribution, confirming some
previous claims (\citealt{Salucci_Burkert00}). These toy-models predict DM slopes which, according to the total mass density slopes, are steepening with stellar mass. At fixed \mst, smaller $r_{\rm B}$ values produce steeper DM slopes. The analysis of this aspect is beyond the scope of this paper, and we will discuss these results more extensively in a future paper.

While an overall steepening of the slope is evident, a flattening and possibly an inversion of the trend seems to emerge in the most massive side,
which is populated by earlier-type and gas-poor systems, resembling what is found in ETGs (\citealt{Tortora+14_DMslope}; see later for a direct comparison). This result is not surprising if we look at the  size-mass trend shown in the top panel of \Fig\ref{fig:Re_fdm_slope_Mstar_LTGs_points}, where the structural properties of these massive galaxies seem different from the other systems in the SPARC sample. This might appear at odds with the recent results of \cite{Posti+19}, who, fitting the rotation curves of the SPARC galaxies, have found that the total stellar-to-halo mass ratio (computed at the virial radius) does not bend at high masses, but continues to increase up to the cosmic baryon fraction in the most massive LTGs (a similar trend was also found in \citealt{Shankar+06}). To analyze this apparent discrepancy, we have implemented the best-fit \Mvir--\mst\ relation found in \cite{Posti+19} in our NFW + baryons toy model, and we show the results in the middle and bottom panels of \Fig\ref{fig:Re_fdm_slope_Mstar_LTGs} (dashed red line).  A different \Mvir--\mst\ relation has a small impact on both \fdm\ and \amw, with only a somewhat less pronounced inversion in the trend of DM fraction with respect to the reference model assuming a \cite{Moster+10} relation. Therefore, this comparison confirms a well known result: the trend between the central \fdm\ and \mst\ is critically dependent firstly on the \Re--\mst\ relation and secondly on the global \Mvir--\mst\ relation  (e.g \citealt{SPIDER-VI}).

All these results are valid if the IMF is the same for all galaxies,
and we do not have any indication that IMF is systematically changing within the SPARC sample. However, performing a dynamical modelling of spiral galaxies from the MANGA survey (mostly at $\mst \gsim 10^{10} \, \Msun$),
\cite{Li+17_IMF} have shown that these galaxies present a similar systematic variation with velocity
dispersion than ETGs, but with a slightly different slope and a larger scatter. Their results may be applicable only for the most massive galaxies
in our sample; we do not have information about IMF variations in dwarf LTGs. In general, the variation of the IMF can alter the trends of
\fdm\ with velocity dispersion, as shown in
\cite{TRN13_SPIDER_IMF} for ETGs, but it is not trivial to
understand the impact in our
\Fig\ref{fig:Re_fdm_slope_Mstar_LTGs}, where \fdm\ and \amw\ are
plotted as a function of \mst. In any case, while the \amw--\mst\
trend is negligibly affected, due to the change in the
\mst\ values only, a variation in IMF more strongly impacts the \fdm\
trend. We cannot include in our analysis a systematic variation of IMF, which has not been clearly determined yet, but we can analyse how \fdm\ changes when the IMF is systematically changed, assuming the values $\Yst = 0.5$ and $0.7
\, \Ysun$. In these cases, the variation of \fdm\ with respect to the reference
value is of $\sim \pm 10$ per cent. However, a more detailed analysis of this
aspect is beyond the scope of this paper.

\begin{figure}
\psfig{file=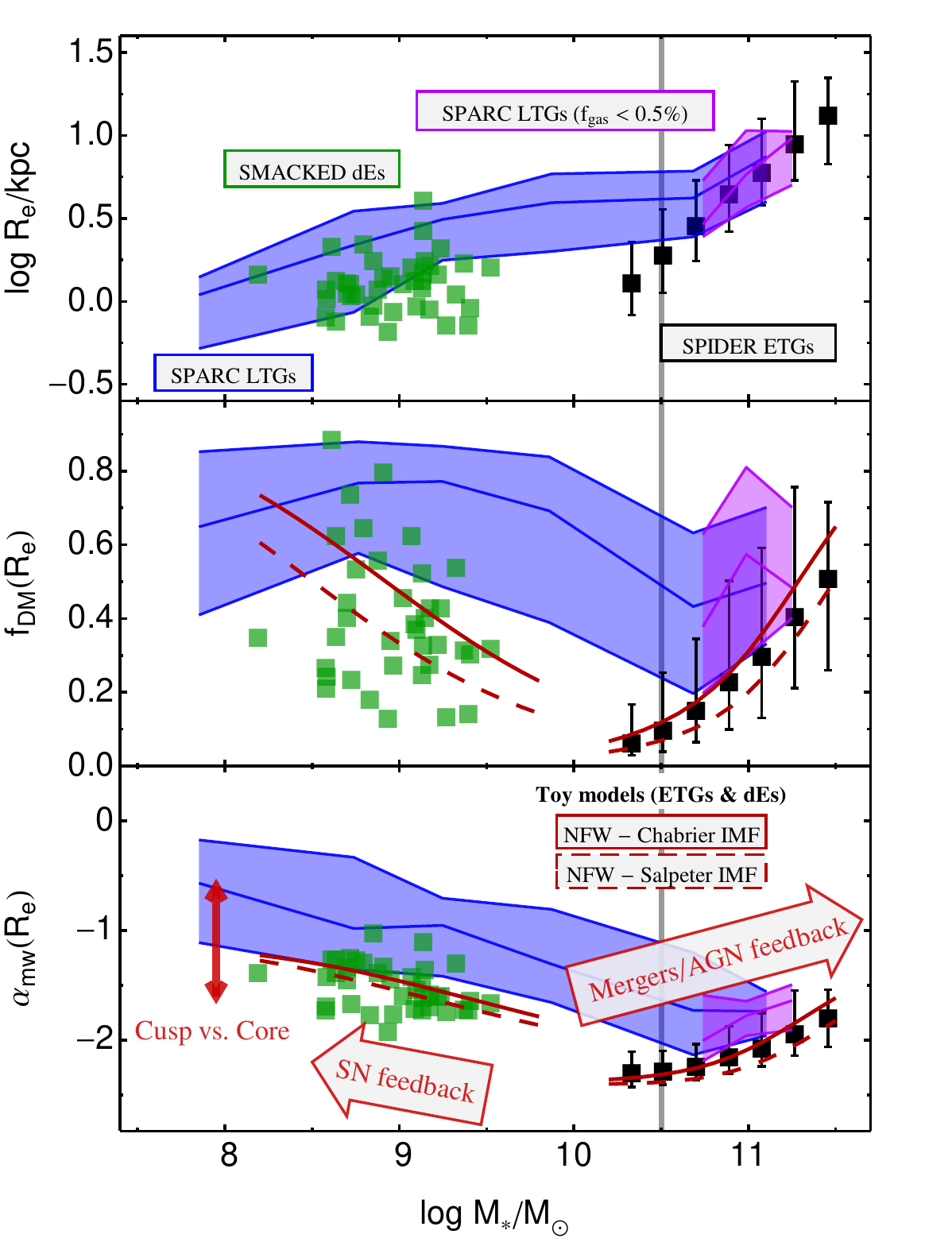, width=0.49\textwidth}
\caption{\Re, \fdm\ and \amw\ as a function of
Chabrier-IMF-based stellar mass for different samples. As in
\Fig\ref{fig:Re_fdm_slope_Mstar_LTGs}, spirals are plotted as blue
lines and shaded regions. Purple lines with shaded regions plot
medians and 16-84th percentiles for LTGs with the lowest amount of
gas (i.e. $f_{\rm gas} < 0.5$ per cent). Black squares with bars are
medians and 16-84th percentiles for SPIDER galaxies, assuming an
NFW profile + baryons and \Ystv\ free.  Green squares are for
SMACKED dEs, assuming NFW + baryons and with \Ystv\ free. Red lines
are toy-models based on our reference NFW model and a S\'ersic
profile with a Chabrier (solid) and Salpeter (dashed) IMF, and
adopting the size-mass relations for dEs and ETGs. The grey vertical line corresponds to the characteristic mass scale of $\sim 5 \times 10^{10} \, \rm \Msun$. The red arrows give information about the phenomena driving the dichotomy and their efficiency with mass. We also add a vertical arrow to point out the difference in mass density slopes among dEs and LTGs, which we relate to a cusp-core transformation in the DM distribution. See the text for more details.}\label{fig:Re_fdm_slope_Mstar_ALL}
\end{figure}

\subsection{Comparison with early-type
galaxies}\label{subsec:comparison_ETGs_dEs}

To complement our study we also include the results for massive
ETGs from the SPIDER survey, which we have worked out in previous
analysis
(\citealt{SPIDER-VI,TRN13_SPIDER_IMF,Tortora+14_DMevol}). To
further study the low-mass regime, we have also considered the
results for dEs using the SMACKED sample
(\citealt{TRN13_SPIDER_IMF}).

In \Fig\ref{fig:Re_fdm_slope_Mstar_ALL}, we start showing the
results for massive ETGs using the reference NFW profile with free
IMF, introduced in \Sec\ref{subsec:SPIDER}. The reason why we consider these results as reference for ETGs is that we have demonstrated that the internal dynamics in ETGs can be realistically described if the IMF is not universal. Otherwise we should recur to unrealistic values of \cvir\ and \Mvir\ of the DM halo. The IMF is found to be ``more massive'', i.e. produces a larger stellar mass, at higher velocity dispersion. However, it is pretty
constant with stellar mass, pointing to a median IMF in between a
Chabrier and a Salpeter IMF shape. We refer the reader to Figure~2 of
\cite{TRN13_SPIDER_IMF} and Figure~1 of \cite{Tortora+14_DMslope},
where these results are found and amply discussed. These results
agree with a plethora of independent works using
different techniques and data samples (e.g.,
\citealt{Treu+10}; \citealt{Conroy_vanDokkum12b};
\citealt{Cappellari+12}; see \citealt{TRN13_SPIDER_IMF} and \citealt{Tortora+14_DMslope} for a comprehensive list of references).

Thus, DM fraction within one \Re\ is
an increasing function of stellar mass, pointing to about $50$ per cent
of DM in the most massive and biggest ETGs with $\mst\sim 3 \times
10^{11}\, \rm \Msun$ and $\Re \sim 10 \, \rm kpc$. On the
contrary, the lowest-mass ETGs are the smallest systems with $\Re
\sim 1 \, \rm kpc$ and with less DM (less than 10 per cent). In the bottom panel, we also show the variation of \amw\ with
mass, which points to steeper mass density profiles at the lowest
masses, approaching the isothermal law at the most massive side
(see \citealt{Tortora+14_DMevol} for more details and results for
other model assumptions). At fixed \mst, ETGs have steeper slopes than LTGs, this is driven by both the steeper stellar-mass density profiles in the former, which have systematically larger S\'ersic indices, and/or steeper DM density profiles.

For consistency with ETG results, we use the same NFW with free IMF model for the dwarf ellipticals,
and add the resulting \fdm\ and \amw\ to the plot (full green
squares). The trends with stellar mass are inverted with respect to the ones
found for the massive ETGs. DM fractions span a wide range of
values and mass density slopes have values in the range $(-2,-1)$.
The lowest mass dEs are expected to have more DM and shallower slopes.

To guide the reading of the trends,
the expectations for the NFW toy-models for two IMF choices are
also overplotted as red lines\footnote{If we consider that the
NFW toy-model is practically the model adopted to derive \fdm\ and
\amw\ with the \Ystv\ free to vary in \Sec\ref{subsec:ETGs_dEs}, it does not surprise the very
good agreement.} (Chabrier IMF with the solid line and Salpeter
IMF with the dashed one). These toy-models have $\alpha_{\rm DM} \sim -1.1$ for LTGs of all stellar masses, which is consistent with the halo fits in \cite{Posti+19}, while for ETGs and dEs these are slightly shallower, but still constant with \mst, since they have smaller \Re\footnote{We caution the reader that a precise comparison of the effective radii for LTGs and ETGs is not trivial, since they are determined with different approaches and in different wavebands.}. In \Fig\ref{fig:Re_fdm_slope_Mstar_ALL} we also show the mass scale where the inversion in the trends is seen, indicating the physical processes which can lead to such different behaviours. We will discuss the physical interpretation of our results in \Sec\ref{sec:phys_interpretation}.

\begin{figure}
\psfig{file=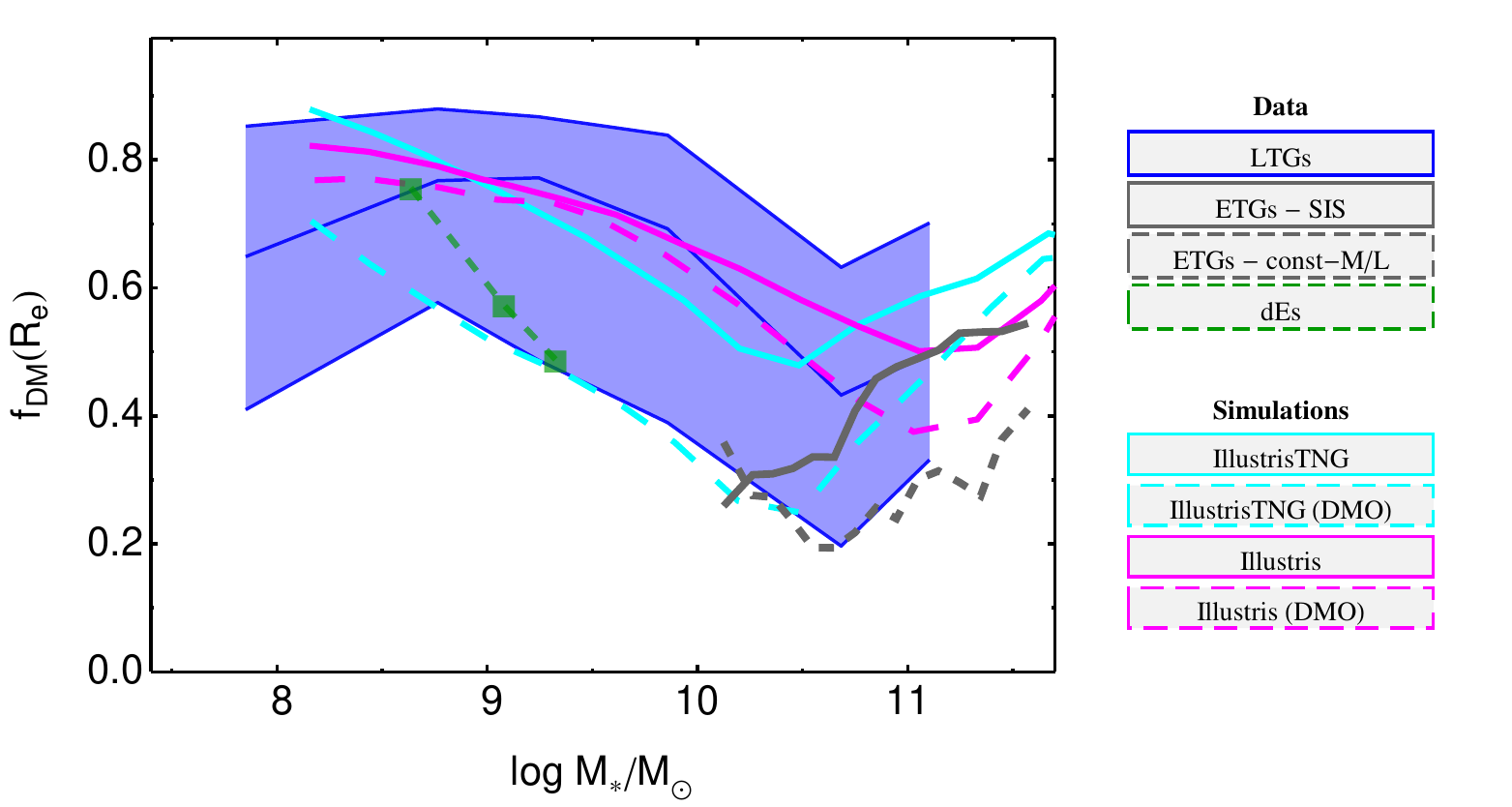, width=0.49\textwidth}
\caption{Comparison with cosmological simulations. We plot
\fdm\ as a function of \mst\ for the different samples analysed
and the outcomes from hydrodynamical simulations. Solid and dashed
grey lines are medians for SPIDER ETG sample, adopting the SIS and
constant-\ML\ models for the total mass profile, respectively, and
using a Chabrier IMF. Dashed green line with squares represent the
median for SMACKED dEs, assuming a constant-\ML\ model for the
total mass profile and a Chabrier IMF. Purple and cyan lines are
for simulated galaxies from Illustris and IllustrisTNG
(\citealt{Lovell+18_Illustris}). Dashed lines are created when the
galaxies, simulated within the full-physics simulations (Illustris
or IllustrisTNG), are placed in their corresponding DM haloes
simulated within the DMO
simulations.}\label{fig:fdm_vs_Mstar_ALL_vs_SIMS}
\end{figure}

\subsection{Dark matter fraction from hydrodynamic
simulations}\label{subsec:comparison_simulations}

Finally, in \Fig\ref{fig:fdm_vs_Mstar_ALL_vs_SIMS} we compare
our \fdm\ with the ones from hydrodynamical simulations in
\citet[Figure~6]{Lovell+18_Illustris}\footnote{Unfortunately,
we have not found similar results in the literature for the total
mass density slope, since most of the works are focussing to small
ranges of masses (typically massive ETGs), adopt variegated mass
density slope definitions and probe different radial scales.}.
Median \fdm\ for the Reference Illustris and IllustrisTNG
simulations are shown as continuous lines. We also show the median
\fdm\ created placing the galaxies, simulated within the
full-physics simulations, in their corresponding DM haloes
simulated within the dark matter only (DMO) simulations. These
latter models neglect the effects of baryonic physics on the DM
distribution. The simulation results assume a universal Chabrier
IMF. They are also calculated within the deprojected half-light
radius, which can be up to 1.6 times larger than the projected
effective radius and the two are equivalent for galaxies with a
stellar mass larger than $\sim 10^{10.5} \, \Msun$
(\citealt{Genel+18}). Therefore, the simulated \fdm\ could be
overestimated at low-masses.

To perform a more homogeneous comparison, we replace our reference
results for dEs and ETGs, with the results assuming a Chabrier IMF
and the two alternative models introduced in \Sec\ref{sec:data}.
Therefore, we adopt the SIS and constant-\ML\ profile for the
SPIDER sample (\citealt{SPIDER-VI}), and the constant-\ML\ profile
for the SMACKED sample (\citealt{TLBN16_IMF_dwarfs}). The results for the reference NFW + baryons model are not shown not to clutter the plot. For massive ETGs, a similar, but more gentle,
variation with mass is found with respect to the reference NFW
model. The SIS produces larger \fdm\ when compared with the
reference NFW + baryons model, especially at low masses. This is
expected since for this latter model a) the IMF is "heavier" than
the Chabrier one and b) the profile is systematically steeper than
$\alpha=2$ at low masses. On average, the steeper constant-\ML\
profile provides \fdm\ more similar to the reference model, but
also an almost constant trend with \mst. For dEs, assuming the
constant--\ML\ profile, we find larger \fdm\ than the reference
model and a steeper trend with mass. This difference is related to
the different IMF and to the higher star formation efficiency in
the Moster relation, which forces \fdm\ to lower values when the
NFW + baryon model is adopted. It is interesting to notice that
the dEs and ETGs reproduce the U-shape trend independently of the
mass model adopted (see \Fig\ref{fig:Re_fdm_slope_Mstar_ALL}).

Except for the TNG100 DMO simulations, the other simulations are
pretty consistent with our findings for LTGs, reproducing the
moderate decline in terms of \mst\ and the inversion of the trend at
large masses. In general, the models are in better agreement with
the most massive ETGs, reproducing both the trend and the
normalisation. This is particularly true if we consider the case
of the isothermal profile in SPIDER ETGs. A good agreement is also
found for the lowest-mass dEs. The full-physics simulations
produce large DM fractions, while the DMO simulations provide
lower DM fractions, which are in better agreement with ETGs.
However, a more homogeneous comparison should be made adopting the
proper projected half-mass radii (or the related light-weighted
values) in the simulations.

\section{Physical interpretation}\label{sec:phys_interpretation}

The results for dEs and ETGs shown in \Fig\ref{fig:Re_fdm_slope_Mstar_ALL} point to a dichotomy of DM content and mass density slope. These results are independently confirmed by Jeans models applied to MANGA galaxies (\citealt{Li+19_MANGA}) and results from hydrodynamic simulations (\citealt{Lovell+18_Illustris}). Larger \fdm\ and shallower slopes are found in the most massive
ETGs ($\mst \gsim 10^{11}\, \rm \Msun$) and the lowest-mass dEs
($\mst \sim 10^{9}\, \rm \Msun$), and a minimum in the DM fraction
and the steepest slopes are seen at the characteristic mass scale
of $\mst \sim 3 \times 10^{10}\, \rm \Msun$. The trends found for
LTGs can, therefore, be compared with these independent results. If
we consider objects with a fixed mass of $\sim 10^{9}\, \rm
\Msun$, then we see that LTGs are more DM dominated, within \Re,
than dEs (see \Fig\ref{fig:Re_fdm_slope_Mstar_LTGs_points}).
LTGs also have shallower total density slopes than dEs of similar
masses. Also, while DM fraction seems to have a more gentle variation
with mass, with a plateau extending till $\mst \sim 10^{10}\, \rm \Msun$,
the steepening of the mass density slope is found to be very similar
in dEs and LTGs with $\mst\lsim 10^{10}\Msun$.

The U-shape behaviour of \fdm\ and \amw\ with \mst\ can be
understood as a result of different feedback mechanisms in these
systems at different mass scales (see \Fig\ref{fig:Re_fdm_slope_Mstar_ALL}). In the lowest mass galaxies
(dEs), star-formation is likely inhibited by (e.g.) supernovae
feedback, which is supposed to be powerful if the potential well
is not too deep, as in these low mass systems. The differences
in slopes observed among dEs and LTGs of similar
mass could be explained by a DM cusp-core transformation
induced by such stellar feedback. Without
recurring to the hypothesis of new physics about DM, within
$\Lambda$CDM framework, simulations tell us that the initial cusps
of DM distributions of dwarf galaxies can be transformed into
cores of size $\sim$ the 3D stellar half-light radius (i.e., of
the same order of magnitude of the projected half-light radius).
Multiple bursts of star formation induce a rapid expansion of the
gas through supernova feedback heating (e.g., \citealt{Pontzen_Governato2012,
Pontzen_Governato2014}; \citealt{Read+16}). The light profiles
of dEs and LTGs of similar masses are not too dissimilar, though
dEs tend to be smaller. Thus, since LTGs are systematically found
to have larger \amw, it is likely that it is so because the DM
distribution is different from that of dEs. Star formation in dEs stopped very early on, in fact they are old and red; while LTGs
of similar mass had a more prolonged star formation history, thus
possibly inducing larger sizes and a transformation of the original central DM density
cusp into an extended core.

Low-mass galaxies, typically high--z LTGs, are built by cold streams, and present a sustained early star formation, which is then regulated by supernova feedback, with an efficiency changing with the mass of the galaxy.
Supernova feedback is supposed to be more efficient in halting
star formation in the lowest mass dEs and LTGs, where the
potential well is not deep. Instead, the deeper potential wells
in more massive galaxies are contrasting this process
(\citealt{dek_birn06}; \citealt{Cattaneo+08}). Therefore galaxies
become more efficient in converting gas into stars, DM fractions
decrease, and the initial cuspy DM distributions are less
efficiently converted in shallower profiles. If not altered
by external (e.g. mergers) or internal violent agents (e.g. AGN
activity), this trend seems to continue up to the highest-mass LTGs
($\mst\sim 10^{11}\Msun$), where the largest star formation
efficiencies are found (\citealt{Posti+19}). However, mergers occurring in the most massive LTGs cannot be excluded. As expected, in our sample, galaxies with a non-zero bulge-to-total mass ratio are typically found at high masses, where both secular evolution, minor and major merging can be responsible for the presence of a bulge (e.g. \citealt{Weinzirl+09_BT_merging}).

For ETGs, which dominate the high-mass end of the galaxy mass
function, the situation is, instead, completely different at
the $\mst\gsim \, 3\times 10^{10}\Msun$, as additional processes, such as dry merging and AGN feedback, play
a fundamental role in inhibiting gas cooling and quenching their
star formation (\citealt{Moster+10}; \citealt{Tortora+10CG}). In fact, the trends
in the total density slope found for massive ETGs can be explained
by dissipation and galaxy merging occurrence. In-Situ star formation,
resulting from dissipative processes, tends to form steeper-than-isothermal
profiles, while gas-poor mergers are a natural attractor towards
the isothermal slope (\citealt{Remus+13,Remus+17}). Thus, in ETGs
with mass $\sim 3 \times 10^{10}\, \rm \Msun$ gas dissipation is
dominant, producing more stars in the cores, smaller effective
radii and \fdm\ and steeper total mass density profiles. Such low-mass ETGs cannot be formed by the merging of LTGs of similar mass, which have larger sizes and shallower slopes (\Fig\ref{fig:Re_fdm_slope_Mstar_ALL}), since such a kind of process would increase the effective radius (\citealt{Naab+09}; \citealt{Hilz+13}), make the density profile shallower (e.g. \citealt{Dehnen+05}) and make the galaxies more DM dominated (\citealt{Tortora+18_KiDS_DMevol}).
In the
most massive ETGs, galaxy (minor) mergers are producing large \Re\ and
\fdm\ (\citealt{Tortora+18_KiDS_DMevol}) and shallower,
approximately isothermal, mass profiles (\citealt{Remus+13,
Remus+17}). Hence, as for the lowest-mass systems, the
highest-mass galaxies are found to have the lowest star-formation
efficiencies, the highest DM content and shallower slopes.

The U-shape trends in \fdm\ and \amw\ for dEs, ETGs, and LTGs add
up to other well-known non-monotonic correlations for galaxies
(see introduction for a list of references).
We found similar differences in terms of galaxy types and mass
in \cite{Tortora+10CG} and \cite{Tortora+11MtoLgrad}, analyzing optical colour and \ML\ gradients in
samples of local dEs, ETGs and LTGs. dEs and ETGs manifest a
similar U-shape trend with stellar mass, with the steepest colour
gradients at $\mst \sim 3 \times 10^{10}\, \rm \Msun$. LTGs have
colour gradients that follow the same steepening with mass found
for dEs, but systematically steeper.

\section{Conclusions}\label{sec:conclusions}

In this paper we have investigated the DM fraction and the total
mass density slopes in the central regions
of late-type galaxies from the SPARC data-sample
(\citealt{Lelli+16_SPARC}), assessing how these quantities vary
with stellar mass. One of the advantages of this
analysis consists in the fact that observed rotation velocities
provide a direct way to calculate both DM fraction and total mass
density profile. While DM fraction can depend on the assumption of
a universal IMF, total mass density slopes are determined without
any mass modelling assumption. We find that the DM fraction is lower
at the highest masses and the mass density profile is shallower in
dwarf LTGs and steeper, approaching the isothermal profile, at the
massive side. We describe these quantities with an approach which
is coherent with previous analyses which were mainly
focussed on the DM fraction and mass density profile in ETGs and
dEs using Jeans equations
(\citealt{SPIDER-VI,TRN13_SPIDER_IMF,Tortora+14_DMslope}).

The trend of DM fraction and mass density slope with stellar mass
has a U-shape behaviour, with largest \fdm\ in most massive ETGs
($\mst \gsim 10^{11}\, \rm \Msun$) and dEs ($\mst \sim 10^{9}\,
\rm \Msun$), and a minimum at $\mst \sim 3 \times 10^{10}\, \rm
\Msun$. At low masses, we have also added the results for LTGs,
which qualitatively resemble the trends with mass found for dEs,
although these latter are spanning a more limited mass range. We
also find that LTGs are more DM dominated and present shallower
mass density slopes than dEs. We suggest that this result can be
explained by a DM cusp-core transformation, induced by stellar
feedback.

All these trends mirror those of the dynamical \ML\
(\citealt{Wolf+10}; \citealt{Toloba+11_I}), and of the total star
formation efficiency with respect to mass and galaxy type
(\citealt{Benson+00}, \citealt{MH02}, \citealt{vdB+07};
\citealt{CW09}; \citealt{Moster+10}; \citealt{Dutton+10};
\citealt{More+11}), as such as the trend of optical colour
gradients with mass (\citealt{Tortora+10CG,Tortora+11MtoLgrad})
which are the result of the interplay among different physical
processes, such as SN feedback at the lowest galaxy masses, and
either AGN feedback and galaxy merging in the most massive
passive galaxies (\citealt{Tortora+10CG}), or an undisturbed and
prolonged star formation activity in massive, star forming spirals
(\citealt{Posti+19}). While in the population of LTGs the global star formation efficiency (\citealt{Posti+19}) and optical colour gradients (\citealt{Tortora+10CG}) seem to be monotonic functions of the stellar mass, at $\mst \gsim 3\times10^{10}\Msun$ ETGs appear to have an opposite trend, being less star forming and having shallower colour gradients as mass increases. However, even when considering spirals only, we see that we cannot exclude a flattening of \fdm\ and \amw\ with stellar mass, since this is mostly driven by a bimodality in the mass-size (which may be due to the structure of discs, e.g. \citealt{TullyVerheijen97}, or to the more frequent presence of bulges in high-mass LTGs).

In the future, we plan to further investigate the properties of
LTGs and their mass density slopes also in terms of the
environment and redshift, discriminating among central and global properties. We plan to improve these estimates also
for dEs, adding more galaxies to the sample, and for ETGs, including
higher-quality and radially extended data, which allows to derive
results which are less dependent on mass modelling
(\citealt{Pulsoni+18_PNS}). Simulations represent a benchmark to interpret the physics
behind the observational results. Defining the DM fraction and
total mass density slope in a homogeneous way for both observations
and simulations is a crucial step to understand the main physical
processes (\citealt{Mukherjee18_SEAGLEI}). We will improve this
aspect using EAGLE simulations, producing mass profiles for
galaxies over the five dex in mass analysed in this paper and
studying their evolution with cosmic time.

%%%%%%%%%%%%%%%%%%%%%%%%%%%%%%%%%%%%%%%%%%%%%%%%%%%%%%%%%%%%%%%%%%%%%%%

\section*{Acknowledgments}
CT and LVEK are supported through an NWO-VICI grant (project
number 639.043.308). CT also acknowledges funding from the INAF
PRIN-SKA 2017 program 1.05.01.88.04. LP acknowledges financial
support from a VICI grant from the Netherlands Organisation
for Scientific Research (NWO) and from the {\it Centre National
d'Etudes Spatiales} (CNES). NRN acknowledges financial support from the one hundred talent program of Sun Yat-sen University and from the European Union Horizon 2020 research and innovation programme under the Marie Skodowska-Curie grant agreement n. 721463 to the SUNDIAL ITN network.

%%%%%%%%%%%%%%%%%%%%%%%%%%%%%%%%%%%%%%%%%%%%%%%%%%%%%%%%%%%%%%%%%%%%%%%

\bibliographystyle{mnras}   % (uses fill "plain.bst")

%\bibliography{Slopes_LTGs_ETGs_FIN}       % expects file "myrefs.bib"

\bibliography{Slopes_LTGs_ETGs_FIN}

\bsp    % typesetting comment
\label{lastpage}
\end{document}